\documentclass[notitlepage,twocolumn,fullpage,showpacs,superscriptaddress,showkeys]{revtex4-1}
\usepackage[utf8]{inputenc}
\usepackage{amsmath}
\usepackage{amsfonts}
\usepackage{amssymb,bbold}
\usepackage{times,fullpage}
\usepackage{comment,float}
\usepackage{tcolorbox,framed}
\usepackage{array,graphicx}
\usepackage[colorlinks=true,citecolor=blue,linkcolor=blue,urlcolor=blue]{hyperref}%
\usepackage{color,cancel}
\usepackage{ulem,bm}

\graphicspath{{./figures/}}

\begin{document}

\newcommand{\nwc}{\newcommand}
\nwc{\vs}{\vspace}
\nwc{\hs}{\hspace}
\nwc{\la}{\langle}
\nwc{\ra}{\rangle}
\nwc{\nn}{\nonumber}
\nwc{\Ra}{\Rightarrow}
\nwc{\wt}{\widetilde}
\nwc{\lw}{\linewidth}
\nwc{\ft}{\frametitle}
\nwc{\ben}{\begin{enumerate}}
\nwc{\een}{\end{enumerate}}
\nwc{\bit}{\begin{itemize}}
\nwc{\eit}{\end{itemize}}
\nwc{\dg}{\dagger}
\nwc{\mA}{\mathcal A}
\nwc{\mD}{\mathcal D}
\nwc{\mB}{\mathcal B}
\nwc{\col}[2]{\textcolor{#1}{#2}}

\nwc{\Tr}[1]{\underset{#1}{\mbox{Tr}}~}
\nwc{\D}[2]{\frac{d #1}{d #2}}
\nwc{\pd}[2]{\frac{\partial #1}{\partial #2}}
\nwc{\ppd}[2]{\frac{\partial^2 #1}{\partial #2^2}}
\nwc{\fd}[2]{\frac{\delta #1}{\delta #2}}
\nwc{\pr}[2]{$K(i_{#1},\alpha_{#1}|i_{#2},\alpha_{#2})$}
\nwc{\av}[1]{$\left< #1\right>$}
\nwc{\alert}[1]{\textcolor{violet}{#1}}

\nwc{\zprl}[3]{Phys. Rev. Lett. ~{\bf #1},~#2~(#3)}
\nwc{\zpre}[3]{Phys. Rev. E ~{\bf #1},~#2~(#3)}
\nwc{\zpra}[3]{Phys. Rev. A ~{\bf #1},~#2~(#3)}
\nwc{\zjsm}[3]{J. Stat. Mech. ~{\bf #1},~#2~(#3)}
\nwc{\zepjb}[3]{Eur. Phys. J. B ~{\bf #1},~#2~(#3)}
\nwc{\zrmp}[3]{Rev. Mod. Phys. ~{\bf #1},~#2~(#3)}
\nwc{\zepl}[3]{Europhys. Lett. ~{\bf #1},~#2~(#3)}
\nwc{\zjsp}[3]{J. Stat. Phys. ~{\bf #1},~#2~(#3)}
\nwc{\zptps}[3]{Prog. Theor. Phys. Suppl. ~{\bf #1},~#2~(#3)}
\nwc{\zpt}[3]{Physics Today ~{\bf #1},~#2~(#3)}
\nwc{\zap}[3]{Adv. Phys. ~{\bf #1},~#2~(#3)}
\nwc{\zjpcm}[3]{J. Phys. Condens. Matter ~{\bf #1},~#2~(#3)}
\nwc{\zjpa}[3]{J. Phys. A ~{\bf #1},~#2~(#3)}
\nwc{\zpjp}[3]{Pramana J. Phys. ~{\bf #1},~#2~(#3)}

\title{Micro Heat Engines With Hydrodynamic Flow}
\author{P. S. Pal}
\email{pspal@kias.re.kr}
\affiliation{School of Physics, Korea Institute for Advanced Study, Seoul 02455, Korea}
\author{Sourabh Lahiri}
\email{sourabhlahiri@bitmesra.ac.in}
\affiliation{Department of Physics, Birla Institute of Technology Mesra, Ranchi 835215, India}
\author{Arnab Saha}
\email{sahaarn@gmail.com}
\affiliation{Department of Physics, University Of Calcutta, 92 Acharya Prafulla Chandra Road, Kolkata-700009, India}

\begin{abstract}
Hydrodynamic flows are often generated in colloidal suspensions. Since colloidal particles are frequently used to construct stochastic heat engines, we study how the hydrodynamic flows influence the output parameters of the engine. We study a single colloidal particle confined in a harmonic trap with time-periodic stiffness that provides the engine protocol, in presence of a steady linear shear flow. The nature of the flow (circular, elliptic or hyperbolic) is externally tunable. At long times, the work done by the flow field is shown to dominate over the thermodynamic (Jarzynski) work done by the trap, if there is an appreciable deviation from the circular flow. The work by the time dependent trap is the sole contributor only for a perfectly circular flow. We also study an extended model, where a microscopic spinning particle (spinor) is tethered close to the colloidal particle, i.e. the working substance of the engine, such that the flow generated by the spinor influences the dynamics of the colloidal particle. We simulate the system and explore the influence of such a flow on the thermodynamics of the engine. We further find that for larger spinning frequencies, the work done by the flow dominates and the system cannot produce thermodynamic work. 

\end{abstract}

\maketitle

\section{Introduction}

Energy harvesting or, energy scavenging is a topical research area where one explores the conversion of ambient energy present in the environment into other forms of energy, such as mechanical energy, electrical energy etc. that can be used to produce thermodynamic work. In this context, together with the advent of advanced technological support, excavating microscopic world has become immensely important due to its potential applications in nano-machinary \cite{browne2006making,balzani2000artificial,abendroth2015controlling,van2007motor}, nanoscale assembly \cite{bishop2009nanoscale,boles2016self}, micro-fluidic \cite{niculescu2021fabrication} and (bio)chemical sensing technologies \cite{grieshaber2008electrochemical} etc. Hence energy harvesting in various forms is now widely attractive in industry and academia.

There is a significant demand of the forefront research on energy harvesting in microscopic world. However there are severe challenges as well. One of the important challenges is the fluctuating force or noise present in the surrounding environment of a microscopic system. The source of the fluctuating forces is thermal when the environment is equilibrated at a temperature $T$. The relevant energy scale associated with the microscopic system of our concern is comparable to the thermal energy scale $k_BT$ ($k_B$ is the Boltzmann constant). Hence the thermal fluctuations can play an important role in the dynamics and thermodynamics of the system. Thermal fluctuations, while causing diffusion, also opposes directional motion (if any) of the system, in accordance with the Fluctuation-Dissipation Theorem \cite{kubo1966fluctuation} . Typically it reduces efficacy of the system when it is used to extract thermodynamic work from its surroundings. However, there are clever set-ups where thermal fluctuation plays a constructive role to enhance the efficiency of energy harvesting \cite{reimann2002introduction,astumian2002brownian}. 

In this work we will focus on a microscopic system where a harmonically trapped single colloidal particle is driven by the time-periodic stiffness to extract thermodynamic work from it. It has been shown experimentally \cite{blickle2012realization,martinez2016brownian,martinez2017colloidal} as well as theoretically \cite{schmiedl2007efficiency,rana2014single} that this tiny set-up can be used as a working substance of a micro-heat engine operating between two heat reservoirs which are working as a source and a sink of heat respectively. Being a small system it is intuitive that thermal fluctuations will have major impact on the stochastic thermodynamics of the system.  For example, due to thermal fluctuations, heat, work and efficiency of such micro-heat engines become stochastic. The distribution of the stochastic efficiency has been shown to be bi-modal with interesting large deviation properties \cite{verley2014unlikely, watanabe2022finite}


Apart from the thermal fluctuations, athermal fluctuations can also be present in the environment surrounding the microscopic system. For example, the system can be surrounded by a suspension of living, motile micro-organisms. They are inherently out-of-equilibrium due to their motility. In this case the system is driven by athermal, non-equilibrium fluctuations originating from the incessant collisions between the self-propelling micro-organisms and the system. Unlike thermally equilibrated suspension, such non-equilibrium {\it{active}} suspension can induce directed motion \cite{di2010bacterial,pietzonka2019autonomous}.  Active suspensions can be used as a non-equilibrium heat bath for a colloidal micro-heat engine set-up. Interestingly, it has been shown that energy can be harvested in the form of thermodynamic work, from an active bath by colloidal micro-heat engines \cite{krishnamurthy2016micrometre,saha2018stochastic,saha2019stochastic}. It has also been shown that the active micro-heat engine can be used to extract more thermodynamic work in comparison to their passive counterpart \cite{krishnamurthy2016micrometre}. Theoretically, the efficiency of active micro-heat engine has been estimated where active processes in the heat bath are modelled by active Ornstein-Uhlenbeck process \cite{saha2019stochastic,kumari2020stochastic}, run-and-tumble process \cite{majumdar2022exactly}, aligning interactions \cite{kumari2024stochastic}, etc. The definitions of stochastic thermodynamic quantities  of such devices is yet to reach a consensus\cite{kumari2024stochastic}.    

As the working substance i.e. the particle here is suspended in a fluid, another important factor in this micro-world energy harvesting processes can be the presence of hydrodynamic flows surrounding the microscopic system. The flow can be generated by an external drive such as shear, pressure or conservative forces \cite{squires2005microfluidics}. It can also be generated by motile micro-organisms, particularly when the suspension is active due to the presence of such organisms \cite{elgeti2015physics,pushkin2013fluid}.  As discussed earlier, the effect of active, athermal fluctuations on the dynamics of the colloidal particle has recently been discussed in the context of micro-heating engines. Apart from this, the flows present in the suspension can also influence the stochastic thermodynamics of the colloidal particle \cite{speck2008role}. In this work we will consider the influence of such flows on the thermodynamics of colloidal micro-heat engine. 

In particular, we will illustrate the effect of the hydrodynamic flow present in the bath, on the thermodynamics of a micro-heat engine \cite{martinez2017colloidal}, where the working substance is a single colloidal particle confined by a harmonic trap with time-periodic stiffness. The temperature of the bath switches between the two values $T_{\rm e}$ and $T_{\rm c}$. Here we will consider $T_{\rm e} > T_{\rm c}$ in accordance with a typical engine set-up. The presence of hydrodynamic flows constitute the primary difference of our model from the earlier works. We will consider different types simple flows present in the bath: (a) Taylor-Couette-like \cite{chandrasekhar2013hydrodynamic} linear shear flow, (b) the flow generated by a microsphere spinning in an otherwise quiescent highly viscous fluid, with a fixed rotational axis \cite{dhont1996introduction}. We do not consider any translational motion of the spinning sphere in (b). 

Setting up Taylor-Couette flow is a common route to introduce shear in a system. The flow around spinning  microsphere, particularly is motivated by experiments with spinning Bacterium \cite{dominick2018rotating}. In the presence of such flows, we will calculate average thermodynamic work for a single, harmonically trapped  colloidal particle in a micro-heat engine set-up. Here we consider simple flows to keep the analysis tractable. In principle, one may consider even more complex as well as realistic flows within the fluid (e.g. instead a single spinning Brownian sphere, one may consider many of them) and estimate its effect on the thermodynamics of the micro-engine. Here we will also assume that the flows we consider  can influence the dynamics of the colloidal particle by advecting it whereas the particle, being very small, cannot affect the surrounding flow. Under this assumption, we first detail the theoretical model for the system (sec.~\ref{sec:model}) and its stochastic thermodynamics (sec.~\ref{sec:stochastic_thermodynamics}) below, where we define the thermodynamic quantities of our interest. Next we discuss the effect of two types of flow fields, namely the shear flow and the spinor flow. In sec~\ref{sec:toy_model}, we provide the analytical and simulation results for micro heat engines with shear flow. In sec~\ref{sec:spinorflow}, we present the results for heat engine with spinor flow. Finally, we conclude in sec~\ref{conclusion}.

\section{Model Description}
\label{sec:model}
We consider a Brownian particle moving in three dimensions (3D) and its position is denoted by $\bm r(t)$ at time $t$. The particle is trapped in a time-dependent potential $U(\bm r,t)$. The particle also advected by a flow field $\bm v(\bm r)$.
In the overdamped limit the equation of motion  of the particle is given by 
\begin{align}
\gamma [\dot{\bm {r}} - \bm v(\bm r)] = -\bm\nabla U(\bm r,t) + \hat{D}.~\bm\eta(t).
\label{eom1}
\end{align}
Here the fluctuating force $\bm\eta(t)$ is assumed to be thermal in nature i.e.,  $\langle \eta_i(t)\rangle =0$, $\langle\eta_i(t^{\prime})\eta_j(t^{\prime\prime})\rangle=\delta_{ij}\delta(t^{\prime}-t^{\prime\prime})$ $\forall i,j\in (x,y,z)$. $\hat D=\sqrt{2D}\mathbb{1}_3=\sqrt{2\gamma K_BT}\mathbb{1}_3$ where $\gamma$ is the coefficient of frictional drag between the particle and the surrounding fluid and $\mathbb{1}_3$ is a $3\times 3$ identity matrix. 
$T$ is the temperature of the fluid. The trapping potential $U(\bm r,t)$ is time-periodic. In the first half of the period it expands and in the second half it contracts. Therefore, in the first half the particle is allowed to explore more and more volume with time (expansion), while in the second half the trend gets reversed (compression). Hence, it acts as a microscopic piston for the trapped Brownian particle. Here, during expansion $T=T_{\rm e}$ and during compression $T=T_{\rm c}<T_{\rm e}$.  Henceforth, the subscripts `$\rm e$' and `$\rm c$' would refer to quantities in the expansion and compression steps, respectively.

We consider 
the trapping potential to be harmonic given by
\begin{align}
    U(\bm r,t)=\frac{1}{2}\bigg[k_x(t)x^2+k_y(t)y^2+k_z(t)z^2\bigg].
    \label{eq:U}
\end{align}
For simplicity, we implement linear expansion and compression protocols:
\begin{align}
k_{i}(t)&=k_{i,\rm e}(t)=k_{i,0}\left(1-\frac{t}{\tau}\right), \hspace{0.3cm} 0\le t < \frac{\tau}{2}\nn\\
&=k_{i,\rm c}(t)=k_{i,0}\frac{t}{\tau}, \hspace{1.5 cm} \frac{\tau}{2}\le t < \tau,
\end{align}
where $\tau$ is the time-period and $k_{i,0}$ is the initial value of the stiffness of trap which is a constant.
%

For the purpose of non-dimensionalization of the variables and parameters, we choose $\sqrt{k_BT/k_0}$ as the characteristic length scale $l_{\rm c}$, where $k_0$ is chosen to be the arithmetic mean of $k_{x,0},~k_{y,0}$ and $k_{z,0}$. 
The characteristic velocity scale readily becomes $l_{\rm c}/t_{\rm c}$ where the characteristic time is $t_{\rm c}=\gamma/k_0$.  
One can now readily rewrite the equation of motion with non-dimensional quantities, and obtain the dimensionless noise strength $D \to D/(k_0^2 l_{\rm c}^2 t_{\rm c})$.  This determines how quick/slow the particle can dissipate. If  the above pre-factor is small the particle will dissipate fast and vice-versa. 
Also, the flow field $\bm{v}$ introduces additional time scales which, in the quasistatic limit, should be much smaller than $\tau$. We will return to this point (see sec. \ref{sec:toy_model}) after we specify the flow fields in the following examples.

\section{Stochastic Thermodynamics}
\label{sec:stochastic_thermodynamics}

Before going  into the details of the results, here we will briefly develop the stochastic thermodynamics and identify the thermodynamically relevant quantities (e.g. heat, work and efficiency) following \cite{Speck2007a}. We denote the internal energy of the system as  $U$, the differential of which can be written as $dU=\bm\nabla U\circ d{\bm r}+\partial_tUdt$, where $\circ$ implies product with Stratonovich convention \cite{gardiner1985handbook}. Considering the relative coordinate $\bm R=\bm r-\bm v(\bm r)t$, we obtain 
\begin{align}
    dU &=\bm{\nabla} U\circ d{\bm R}+(\bm v\circ \bm\nabla U) dt + \partial_t U dt\nonumber\\
    &=  \bm\nabla U\circ d{\bm R}+(\partial_t + \bm v\circ\bm\nabla)Udt\nonumber\\
    &= \bm\nabla U\circ d{\bm R}+D_tUdt, 
    \label{eq:dU}
\end{align}
 where $D_t\equiv \partial_t+\bm v\circ \bm\nabla$ is the convective derivative, i.e., the derivative computed by an observer moving along with the flow, but expressed in terms of the coordinates of a \textit{stationary} observer. From this expression of $dU$ and from Eq. \eqref{eom1} one can write the differential heat exchange between the particle and the surrounding fluid along a single trajectory as, 
 \begin{align}
     dQ &\equiv  (\gamma\dot{\bm R}-\sqrt{D}{\bm\eta})\circ\dot{\bm R}dt\nonumber\\
     &= - (dU - D_t U dt)\equiv -dU + dW.
     \label{1stLaw}
 \end{align}
It is the first law of stochastic thermodynamics for the systems with flow \cite{Speck2007a} where the trajectory dependent work is identified as, 
\begin{eqnarray}
dW\equiv D_tUdt=\partial_tUdt+\bm{v}\circ\nabla Udt.
\label{work}
\end{eqnarray}
Note that from Eq.[\ref{work}], the infinitesimal work is a sum of  the  infinitesimal works due to the rate of the change of the potential $U$ (i.e. $\partial_tUdt \equiv dW_{\text{JE}}$) and due to the flow (i.e. $\bm{v}\circ\bm\nabla U dt=dW_{\text{flow}}$). Integrating these infinitesimal works over a cycle and taking averages over all possible realizations, average work due to the time-variation of the potential ($\langle W_{\text{JE}}\rangle$) and due to the flow ($\langle W_{\text{flow}}\rangle$) in a cycle can be obtained.  Clearly, the total average work in a cycle $\langle W\rangle=\langle W_{\text{JE}}\rangle+\langle W_{\text{flow}}\rangle$.

We will calculate stochastic work $W$ done on the engine over a complete cycle by integrating Eq. \eqref{work} along a trajectory. According to the first law, the heat dissipated is given by $Q = W-\Delta U$, where $\Delta U$ is the change in internal energy.  $\langle Q\rangle$ and $\langle W\rangle$ are the values of heat and work averaged over an ensemble of trajectories. According to our convention, when the average work is negative i.e. $\langle W\rangle < 0$, it implies that work is being extracted from the system, else it is being done on the system. When the average work is being extracted (i.e., when the system behaves as an engine), one can define average efficiency as 
\begin{align}
\xi =\frac{|\langle W\rangle |}{|\langle Q_{\text{exp}}\rangle|},
\label{efficiency}
\end{align}
where $Q_{\text{exp}}$ is the heat absorbed on average from the hot bath during the expansion step.
Note that since the convective derivative replaces the ordinary derivatives in the definitions of work and heat, one has effectively averted the problem of divergences of average thermodynamic quantities with growing cycle time, which are otherwise expected on the grounds of having to maintain the flow externally, throughout the time of observation.


\section{Micro-heat engine with shear flow}\label{sec:toy_model}

\begin{figure*}[t]
	\centering
		\includegraphics[width=\textwidth]{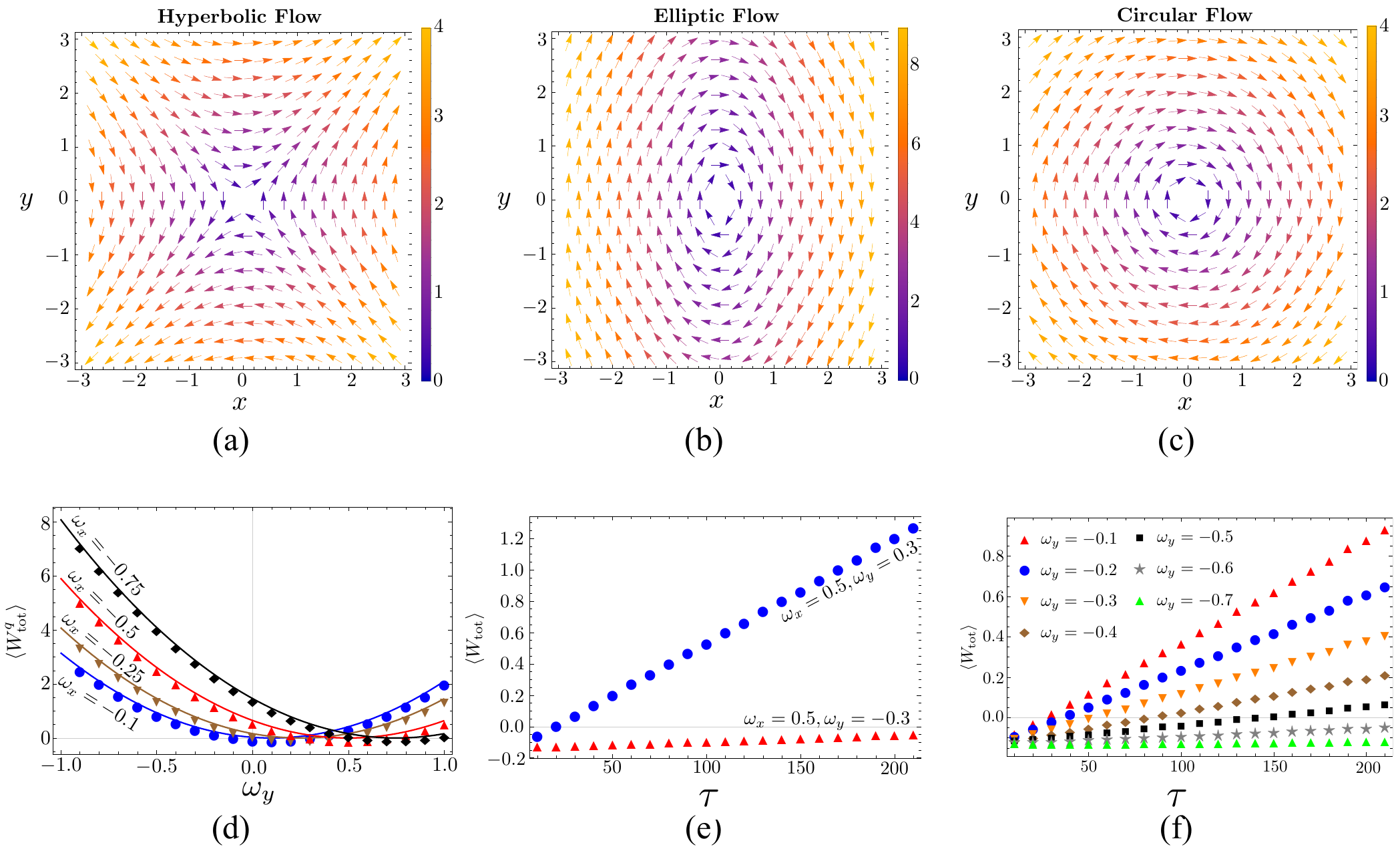}
		\caption{
  The flow field $\bm v=\omega_x y\hat i+\omega_y x\hat j$ is shown. The direction of the flow field is shown by the arrow whereas the magnitude of the flow field is denoted by the color bar. There can be three possibilities depending on the values and signs of $\omega_x$ and $\omega_y$: (a) hyperbolic flow ($\text{sgn}(\omega_x) = \text{sgn}(\omega_y)$) and (b) elliptic flow ($\text{sgn}(\omega_x) = -\text{sgn}(\omega_y)$with $|\omega_x|\neq|\omega_y|$), (c) circular flow ($\omega_x=-\omega_y$).
  (d) Total average work done on the system as a function of one flow parameter $\omega_y$ in the quasistatic limit. The solid lines are obtained from plotting the result in Eq. \ref{q_tot_work} and the different types of dots are obtained by simulating the system at large cycle time ($\tau=250$).
  (e) Total average work done on the system as a function of cycle time for two different sets flow parameters $(\omega_x,\omega_y)$.
  (f) Total average work done on the system as a function of cycle time. The value of $\omega_x$ is fixed at $0.8$. 
  The temperature of the hot and cold bath are $0.2$ and $0.1$ respectively. 
  }
	\label{fig:toy_model_results1}
\end{figure*}

Here we consider that the colloidal particle is bounded in a 2-D plane and is driven out of equilibrium by a breathing isotropic harmonic trap $U(\bm r,t)=\frac{1}{2}k(t)[x^2+y^2]$. The particle is driven by a two dimensional linear shear flow \cite{dhont1996introduction} simultaneously. The flow is represented by  
\begin{equation}
    \bm v =  (\omega_x y)\hat i+(\omega_y x)\hat j.
\end{equation}
The flow couples the $x$ and $y$-coordinates of the Brownian particle, with $\hat i$ and $\hat j$ representing the unit vectors along $x$ and $y$-axes, respectively. Here $(\omega_x,\omega_y)$ are the shear rates (inverse time scales) related to the 2D flow field. They determine the flow geometry. Depending on their values and sign the flow can be circular, elliptic and hyperbolic. In Fig.~\ref{fig:toy_model_results1} different flow geometries are plotted for different values of shear rates. Hence, tuning the shear rates externally, one may tune the geometry of the flow.  Our aim here is to explore how the thermodynamics of the system depends on the flow geometry in the context of the colloidal micro heat engine. One may note here that the special case where $\omega_x=-\omega_y$ can be obtained simply by rotating an incompressible fluid in a cylindrical geometry with no-slip boundary condition, which is commonly known as circular Couette flow \cite{landau1987fluid}. Depending on the shear rates and the strength of the harmonic trap, the trajectory of the particle can become unstable. However, here we consider only the stable trajectories. 



 
In the expansion (compression) step, the system is in contact with the heat bath at temperature $T_{\rm e}$ ($T_{\rm c}$).
The dynamical equation describing the system is
\begin{align}
\frac{d\bm r}{dt}=-\hat\Gamma_{\rm \alpha}.\bm r+\hat D_{\rm \alpha}.\bm \eta,
\label{eq:ceom_exp_mat}
\end{align}
where
\begin{align}
\bm r &=\begin{bmatrix} x & y\end{bmatrix}^T, \hspace{0.2cm}
\bm \eta =\begin{bmatrix} \eta_x & \eta_y\end{bmatrix}^T,\nonumber\\
\hat D_{\rm \alpha} &=\frac{\sqrt{2D_{\rm \alpha}}}{\gamma}\mathbb{1}_2, \hspace{0.2cm}\text{and} \hspace{0.2cm}
 \hat\Gamma_{\rm \alpha} =\frac{1}{\gamma}\begin{bmatrix}k_{\rm \alpha}(t) & -\gamma\omega_x\\-\gamma\omega_y & k_{\rm \alpha}(t)\end{bmatrix},
 \label{eq:MatricesUsed}
\end{align}
with $D_{\rm \alpha}=\gamma k_BT_{\rm \alpha}$ and $\mathbb{1}_2$ is a $2\times 2$ identity matrix. The subscript $\alpha$ corresponds to subscripts `$\rm e$' or `$\rm c$' which denote the expansion or compression steps respectively. The Langevin equation of motions \eqref{eq:ceom_exp_mat} are coupled. In order to proceed, we to transform to a new set of coordinates 
\begin{align}
    \bm r' &=
    \begin{bmatrix} \frac{1}{2}(-\omega ~x+y) & \frac{1}{2}(\omega~x+y)\end{bmatrix}^T \nonumber\\
    &\equiv\begin{bmatrix} x' & y'\end{bmatrix}^T,
\end{align}
in which $\hat\Gamma_{\rm e}$ becomes diagonal. Here $\omega=\sqrt{\omega_x/\omega_y}$.
%
%
In this primed frame of reference  the decoupled Langevin equations can be written as:
\begin{align}
\frac{d \bm r'(t)}{dt} &=-\hat\Gamma'_{\rm \alpha}(t).\bm r'+\hat D'_{\rm \alpha}.\bm \eta',
\label{deom_exp_mat}
\end{align}
where,
\begin{align}
\hat D'_{\rm \alpha}=\hat D_{\rm \alpha}, 
&\hspace{.5 cm}\hat\Gamma'_{\rm \alpha}(t)
=\begin{bmatrix}\lambda^{\rm \alpha}_x(t)&0\\0&\lambda^{\rm \alpha}_y(t) \end{bmatrix}, \hspace{0.2cm}\text{and}\and \nonumber\\ 
\bm\eta'(t)&=\frac{1}{2}\begin{bmatrix}-\omega^{-1}\eta_x(t)+\eta_y(t)\\ \omega^{-1}\eta_x(t)+\eta_y(t)\end{bmatrix}.
\label{eq:PrimedVariablesExpansion}
\end{align}
The time dependent functions (matrix elements) appearing in $\hat\Gamma'_{\rm \alpha}(t)$ are further given by the expressions $\lambda^{\rm \alpha}_x(t) \equiv k_{\rm \alpha}(t)+\gamma\sqrt{\omega_x\omega_y}$, and $\lambda^{\rm \alpha}_y(t) \equiv k_{\rm \alpha}(t)-\gamma\sqrt{\omega_x\omega_y}$. The mathematical steps are detailed in Appendix \ref{sec:decoupling}.

In the primed frame one can now get $\bm r'(t)$ by solving Eq. \eqref{deom_exp_mat}. The general solution for any cycle time is given in the Appendix [See equations: \eqref{sol_comp_x}, \eqref{sol_comp_y}]. From the solutions one can calculate the second moments $\sigma_{x^{\prime}}\equiv \langle x^{\prime 2}\rangle$ , $\sigma_{y^{\prime}}\equiv \langle y^{\prime 2}\rangle$ and $\sigma_{x^{\prime}y^{\prime}}\equiv \langle x^{\prime}y^{\prime}\rangle=\langle y^{\prime}x^{\prime}\rangle$ .  The expressions for the second moments in the primed frame are given in the Appendix (see Eqs. \eqref{eq:variance expansion} and \eqref{eq:variance compression}). These moments are necessary to calculate noise-averaged thermodynamic quantities such as work, heat etc. for an arbitrary cycle time. Next, we focus on the quasistatic regime (large cycle time).

\subsection{Analytical results in the quasistatic regime}

In the limit of a large cycle time (compared to the relaxation time scale), we take the following simple route to calculate the second moments in the primed frame.
Defining the vector  $\bm\sigma_{\rm \alpha}'=\begin{bmatrix}\sigma^{\rm \alpha}_{x'}& \sigma^{\rm \alpha}_{y'} &\sigma^{\rm \alpha}_{x'y'}\end{bmatrix}^T$,  the noise-averaged dynamical equation for $\bm\sigma^{\prime}_{\rm \alpha}$  can be derived from Eq. \eqref{deom_exp_mat} and is given by
\begin{align}
\frac{d\bm\sigma_{\rm \alpha}'}{dt}=-\hat \Lambda_{\rm \alpha}.\bm\sigma_{\rm \alpha}'+\frac{D_{\rm \alpha}}{\gamma^2}\hat\Theta.
\label{2nd_moment_exp}
\end{align}
Here,
\begin{align}
\hat \Lambda_{\rm \alpha}&=\begin{bmatrix}2\lambda^{\rm \alpha}_x(t)&0&0\\0&2\lambda^{\rm \alpha}_y(t)&0\\0&0&\lambda^{\rm \alpha}_x(t)+\lambda^{\rm \alpha}_y(t)\end{bmatrix}\nn\\
\hat \Theta&=\begin{bmatrix}1+\omega^{-2}\\1+\omega^{-2}\\1-\omega^{-2}\end{bmatrix}.\nn
\end{align}
To derive Eq. \eqref{2nd_moment_exp} from Eq. \eqref{deom_exp_mat} we took an inner product of Eq. \eqref{deom_exp_mat}] with $\bm r'(t)$ and then calculated the average $\la \bm r'(t).\hat D'_{\rm \alpha}\bm \eta'(t) \ra$ by using the formal solution of  Eq. \eqref{deom_exp_mat} together with the noise statistics. 
The detailed expressions are provided in the Appendix \ref{sec:SecondMoments}. In the quasistatic limit, the rate of change of $\bm\sigma_{\rm \alpha}'$ becomes negligibly small in comparison to the other term on the right-hand-side of Eq. \eqref{2nd_moment_exp}. Therefore, in quasistatic regime one can simply drop the time derivative of $\bm \sigma_{\rm \alpha}'$ and obtain the quasistatic expression for $\bm\sigma_{\rm \alpha}'(t)$. The expressions are given in Eqs. \eqref{eq:sigma_h primes} and \eqref{eq:sigma_c primes} of the Appendix~\ref{sec:AverageWorkCalculation}. Thus we obtain the expressions of second moments in primed frame of reference in quasistatic limit. 

As the transformation relation between $\bm r(t)$ and $\bm r'(t)$ are linear, the second moments in the primed and unprimed frames also linearly related. Exploiting this fact, the rate of change of mean thermodynamic work can be obtained as discussed below.
%
%
Using the definition of the convective derivative, it is easy to see that (see discussion below Eq. \eqref{eq:dU}) $D_t U(x,y,t) = \partial_t U + \omega_x y\partial_x U+\omega_y x\partial_y U$. In  terms of the coordinates in the \textit{primed frame}, the average rate of work in the \textit{unprimed frame} in the expansion step ($0\le t\le \tau/2$) becomes (see Appendix \ref{sec:AverageWorkCalculation})
\begin{align}
    \la\dot w_{\rm e}(t)\ra&=\la D_tU(x,y,t)\ra \nn\\
    &=-\frac{k_0}{2\tau}\left(1+\frac{\omega_x}{\omega_y}\right)\sigma^{\rm e}_+(t)\nn\\
&\hspace{.35 cm}-\frac{k_0}{\tau}\left(1-\frac{\omega_x}{\omega_y}\right)\sigma^{\rm e}_{x'y'}(t)\nn\\
&\hspace{.35 cm}-(\omega_x+\omega_y)\sqrt{\frac{\omega_x}{\omega_y}}k_0\left(1-\frac{t}{\tau}\right)\sigma^{\rm e}_-(t),
\label{wdot_exp_main}
\end{align}
where $\sigma^{\rm e}_+(t)=\sigma^{\rm e}_{x'}(t)+\sigma^{\rm e}_{y'}(t)$ and $\sigma^{\rm e}_-(t)=\sigma^{\rm e}_{x'}(t)-\sigma^{\rm e}_{y'}(t)$. Using First Law given in Eq. \eqref{1stLaw}, the rate of heat exchanged between the system and the heat bath can also be readily obtained:
\begin{align}
\la\dot q_{\rm e}(t)\ra&=\la\dot w_{\rm e}(t)\ra-\left\la\frac{dU}{dt}\right\ra\nn\\
&=\frac{k^2_{\rm e}(t)}{\gamma}\bigg(1+\frac{\omega_x}{\omega_y}\bigg)\sigma^{\rm e}_+(t)\nn\\
&\hspace{.35 cm}+\frac{2k^2_{\rm e}(t)}{\gamma}\bigg(1-\frac{\omega_x}{\omega_y}\bigg)\sigma^{\rm e}_{x'y'}(t)-\frac{4D_{\rm e}}{\gamma^2}k_{\rm e}(t).
\label{qdot_exp1}
\end{align}

In the compression step ($\tau/2\le t\le \tau$), the system is in contact with cold bath at temperature $T_{\rm c}$.
Using the definition of the convective derivative in the expression of work (see Eq.~\eqref{work}), we obtain the average rate of work done in the compression step:
\begin{align}
\la\dot w_{\rm c}(t)\ra 
&=\frac{k_0}{2\tau}\left(1+\frac{\omega_x}{\omega_y}\right)\sigma^{\rm c}_+(t)
+\frac{k_0}{\tau}\left(1-\frac{\omega_x}{\omega_y}\right)\sigma^{\rm c}_{x'y'}(t)\nn\\
&\hspace{.35 cm}-(\omega_x+\omega_y)\sqrt{\frac{\omega_x}{\omega_y}}\frac{k_0t}{\tau}\sigma^{\rm c}_-(t),
\label{wdot_comp_main}
\end{align}
where $\sigma^{\rm c}_+(t)=\sigma^{\rm c}_{x'}(t)+\sigma^{\rm c}_{y'}(t)$ and $\sigma^{\rm c}_-(t)=\sigma^{\rm c}_{x'}(t)-\sigma^{\rm c}_{y'}(t)$. The average rate of heat exchanged with the cold bath in the compression step becomes 
\begin{align}
\la\dot q_{\rm c}(t)\ra&=\frac{k^2_{\rm c}(t)}{\gamma}\bigg(1+\frac{\omega_x}{\omega_y}\bigg)\sigma^c_+(t)\nn\\
&\hspace{.35 cm}+\frac{2k^2_{\rm c}(t)}{\gamma}\bigg(1-\frac{\omega_x}{\omega_y}\bigg)\sigma^h_{x'y'}(t)-\frac{4D_h}{\gamma^2}k_{\rm c}(t).
\label{qdot_comp}
\end{align}
The quasistatic calculation sets a benchmark for the numerical analysis. The quasistatic limit is obtained by $\tau\rightarrow\infty$ while $t/\tau$ is finite. We consider a new variable (a rescaled time variable) $s=t/\tau$. This implies that $\D{}{t}=\D{s}{t}\D{}{s}=\frac{1}{\tau}\D{}{s}$. Therefore, changing the variable in Eq. \eqref{2nd_moment_exp} and making similar changes for the compression step, after ignoring the terms $\sim O(1/\tau)$ we get
\begin{align}
    \bm\sigma_{\rm e}'=\frac{D_{\rm e}}{2\omega_x\gamma^2}\hat \Lambda_{\rm e}^{-1}.\hat\Theta,\hspace{0.2 cm}
    \bm\sigma_{\rm c}'=\frac{D_{\rm c}}{2\omega_x\gamma^2}\hat \Lambda_{\rm c}^{-1}.\hat\Theta
\end{align}
Using the above expressions in Eqs. \eqref{wdot_exp_main} and \eqref{wdot_comp_main},  the expression of average rate of work in the expansion and compression steps in quasistatic limit (denoted by $q$ as superscript of the corresponding thermodynamic quantities) are respectively (see Appendix \ref{sec:AverageWorkCalculation} )
\begin{align}
\left\la\frac{d w^q_{\rm e}(s)}{ds}\right\ra&=\tau D_{\rm e}(\omega_x+\omega_y)^2\frac{k_0(1-s)}{2[k_0^2(1-s)^2-\gamma^2\omega_x\omega_y]}, \\
\left\la\frac{d w^q_{\rm c}(s)}{ds}\right\ra&=\tau D_{\rm c}(\omega_x+\omega_y)^2\frac{k_0s}{2[k_0^2s^2-\gamma^2\omega_x\omega_y]}.
\end{align}
Note that terms that are $\sim O(1)$ have been dropped in favour of the ones that are $\sim O(\tau)$, since we are considering the quasistatic limit, $\tau\to\infty$. The exact expressions are provided in Eqs. \eqref{eq:WexpExact} and \eqref{eq:WcomExact} in the supplementary materials. The terms which are neglected correspond to $W_{\text{JE}}$ (see Eqs. \eqref{wdot_exp} and \eqref{wdot_comp}). The remaining part, which is proportional to the cycle time $\tau$, is the contribution coming from the energy injected into the system by means of the flow field $\bm v(\bm r)$. 
The total average work in the quasistatic limit is then readily obtained, under the above approximation, by summing up the contributions from the expansion and the compression steps:
\begin{align}
\la W^q_{\text{tot}}\ra&=\int_0^1\left\la\frac{d w^q_{\text{tot}}(s)}{ds}\right\ra ds\nn\\
&=\int_0^{1/2}\left\la\frac{d w^q_{\rm e}(s)}{ds}\right\ra ds+\int_{1/2}^1\left\la\frac{d w^q_{\rm c}(s)}{ds}\right\ra ds\nn\\
&=\frac{\tau (D_{\rm e}+D_{\rm c})(\omega_x+\omega_y)^2}{4k_0}\ln\bigg[\frac{4(k_0^2-\gamma^2\omega_x\omega_y)}{k_0^2-4\gamma^2\omega_x\omega_y}\bigg].
\label{q_tot_work}
\end{align}
From above expression it is clear that in the quasistatic limit , $\langle W^q_{\text{tot}}\rangle>0$. Therefore, in the presence of the flow, irrespective of the values of $\omega_x$ and $\omega_y$, thermodynamic work cannot be extracted in the quasistatic limit. However in the \textit{non-quasistatic} limit one can extract work depending on the values of the flow parameters.  This can be shown by means of simulations. We provide the detailed results in the next section.

\subsection{Simulation results}
\label{sec:ToyModelSimulat}
In Fig. \ref{fig:toy_model_results1}, we have shown the results of our simulations for the flow pattern (subfigures (a)-(c)) and the mean work (subfigures (d)-(f)). In the top panel, the circular flow pattern (subfigure (a), $\omega_x=-\omega_y$ in the flow field $\bm v = \omega_x y \hat i + \omega_y x \hat j$), the hyperbolic flow pattern (subfigure (b), $\text{sgn}(\omega_x) = \text{sgn}(\omega_y)$) and the elliptic flow pattern (subfigure (c), $\text{sgn}(\omega_x) = -\text{sgn}(\omega_y)$, $|\omega_x|\ne |\omega_y|$)  have been shown. Clearly, the elliptic flow field has a higher magnitude of $\bm v$ as compared to the hyperbolic flow for the same magnitudes of $\omega_x$ and $\omega_y$.

In the bottom panel, subfigure (d) provides the variations in the mean quasistatic total work, $\langle W^q_{\rm tot}\rangle$ as a function of $\omega_x$, with $\omega_y$ being used as a parameter. The results of our simulation are shown by points, while the solid lines are the analytical results given by Eq. \eqref{q_tot_work}. 
Since the values of $\omega_x$ used in this subfigure are negative, the range $\omega_y<0$ corresponds to hyperbolic flows, while the range $\omega_y>0$ yields elliptic flows. 
The curves in the elliptic flow range converge to circular flows when the condition $\omega_y=-\omega_x$ is satisfied. For instance, the curve corresponding to $\omega_x=-0.1$ (blue solid line and solid circles) becomes circular at $\omega_y=+0.1$. From Eq. \eqref{q_tot_work} we would then expect the mean total work to vanish at this point. 

However the total work obtained from simulation differs from zero. The difference between the analytical results and the simulations, though small, is important. It occurs from the fact that in the analytical derivation of the average work, we have neglected the terms corresponds to $W_{\text{JE}}$ as it is usually very small in comparison to the work done by the hydrodynamic flow $W_{\text{flow}}$in the quasistatic limit. This is because $W_{\text{flow}}$ increases with the cycle time $\tau $  (which is very large in the quasistatic limit) whereas $W_{\text{JE}}$ is independent of  $\tau$. However, circular flow being the special case with $\omega_x=-\omega_y$ , even in the quasistatic limit the work done by the flow becomes zero (see Eq.[\ref{q_tot_work}]). Therefore in this regime, only $W_{\text{JE}}$ contributes to the total work and produces the difference between the analytical and the numerical results. Clearly, in this regime, even in the presence of hydrodynamic flow, work can be extracted from the system in the quasistatic limit (as $\langle W_{\text{JE}}\rangle <0$). The work extraction in the quasistatic limit is possible even for slightly elliptical flows, as long as $\langle W_{JE}\rangle$ is larger than the work done by the flow.

Nevertheless, since the convergence to circular flows take place at higher values of $\omega_y$ for higher values of $|\omega_x|$, the minima of the curves are observed to shift rightwards as $|\omega_x|$ increases. In general, the work done is found to be higher for hyperbolic flows as compared to elliptic flows.

Subfigure (e) shows the variation in the total work $\langle W_{\rm tot} \rangle$ with the variation in the cycle time $\tau$. The blue solid circles are the simulated results for the above dependence when the flow is hyperbolic ($\omega_x=0.5,~\omega_y=0.3$), while the red solid triangles are those when the flow is elliptic ($\omega_x=0.5,~\omega_y=-0.3$). Consistent with our observations in subfigure (d), in the presence of the hyperbolic flow, a higher amount of work has to be injected into the system. In contrast, the elliptic flow retains the system in the engine mode (work is negative, i.e. extracted from the system) throughout the range of cycle times considered here.

In subfigure (f), we fix the value of $\omega_x$ at 0.8 and show the dependence of $\langle W_{\rm tot}\rangle$ with the cycle time for various values of $\omega_y$, all being in the elliptic flow regime. The weakly elliptic flows ($|\omega_y-\omega_x| \ll |\omega_x|,|\omega_y|$) yield extractable work in the highly non-quasistatic regime (smaller cycle time), while they tend to leave the engine mode (no work can be extracted) in the quasistatic regime. This is observed for the curves for which $|\omega_y|\le 0.5$, thereby showing appreciable deviation from a circular flow, and allowing the flow work to become much larger than the Jarzynski work. On the other hand, the flows which are closer  the circular pattern ( e.g.  $\omega_y=-0.6$ and $-0.7$) can sustain the engine mode (work can be extracted) even for much larger values of $\tau$.




%
%

\begin{figure}[!h]
	\centering
		\includegraphics[width=0.48\textwidth]{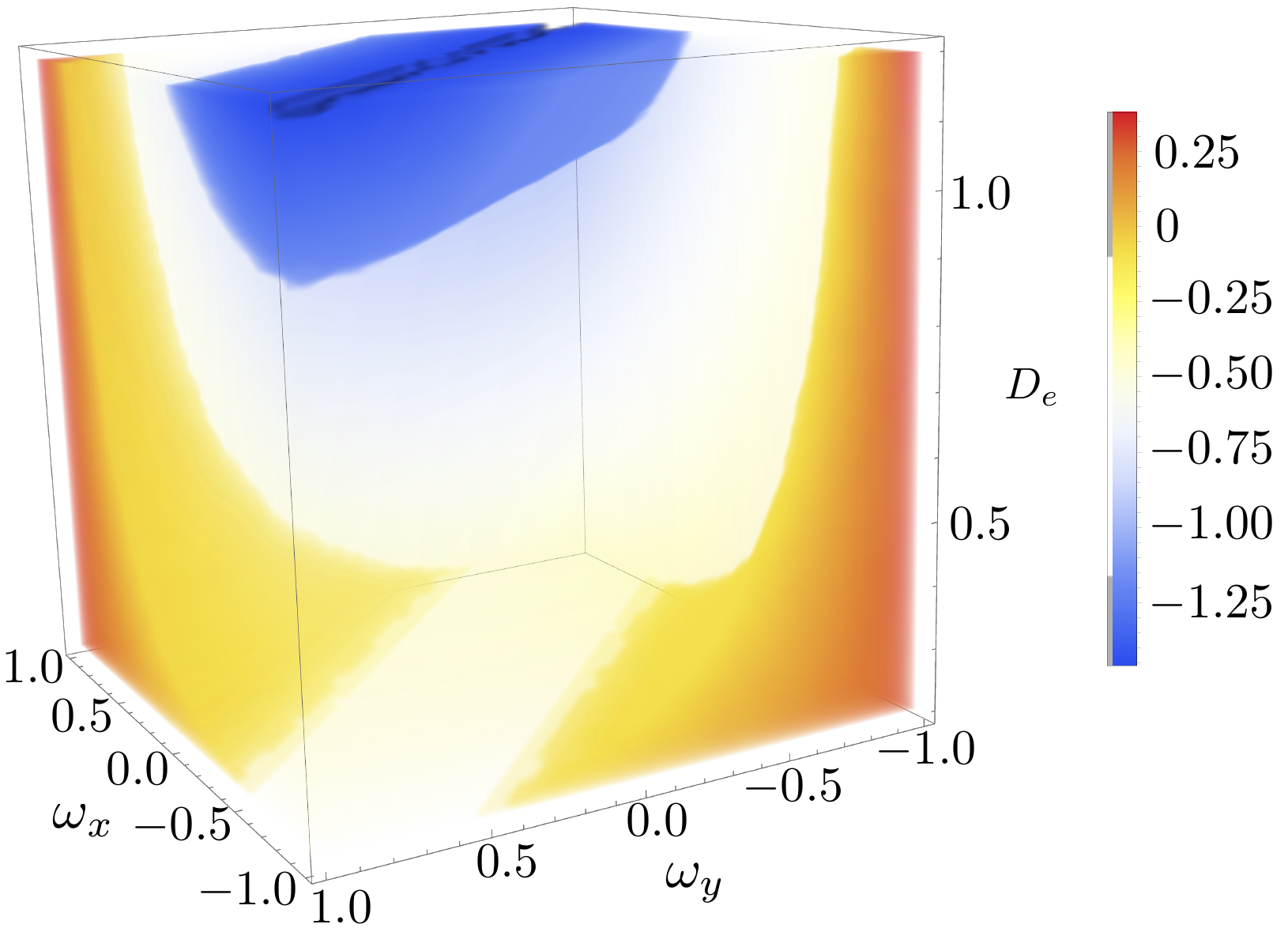}
		\caption{Total average work done on the system as a function of flow parameters $(\omega_x,\omega_y)$ and the hot bath temperature $D_{\rm e}$. Here the cycle time $\tau=10$ and the cold bath temperature $D_{\rm c}=0.1$}
	\label{fig:wtot_vs_om1_om2}
\end{figure}

In the three-dimensional plot of Fig. \ref{fig:wtot_vs_om1_om2}, the variation of $\langle W_{\rm tot}\rangle$ on the values of $\omega_x$ and $\omega_y$ have been shown. The blue region is where the device works as an engine and the work can be extracted maximally within the corresponding ranges of $\omega_x$ and $\omega_y$. Note that this region is a band formed around the diagonal described by $\omega_x=-\omega_y$, again corroborating the fact that circular flows yield more work as output. The hyperbolic flows are found to be less useful for the device to work as engine. They move the system out of the engine mode for higher values of the flow parameters (regions shaded in red color).

\section{Micro-heat engine with Spinor Flow}\label{sec:spinorflow}

\begin{figure}[!h]
	\centering
		\includegraphics[width=0.48\textwidth]{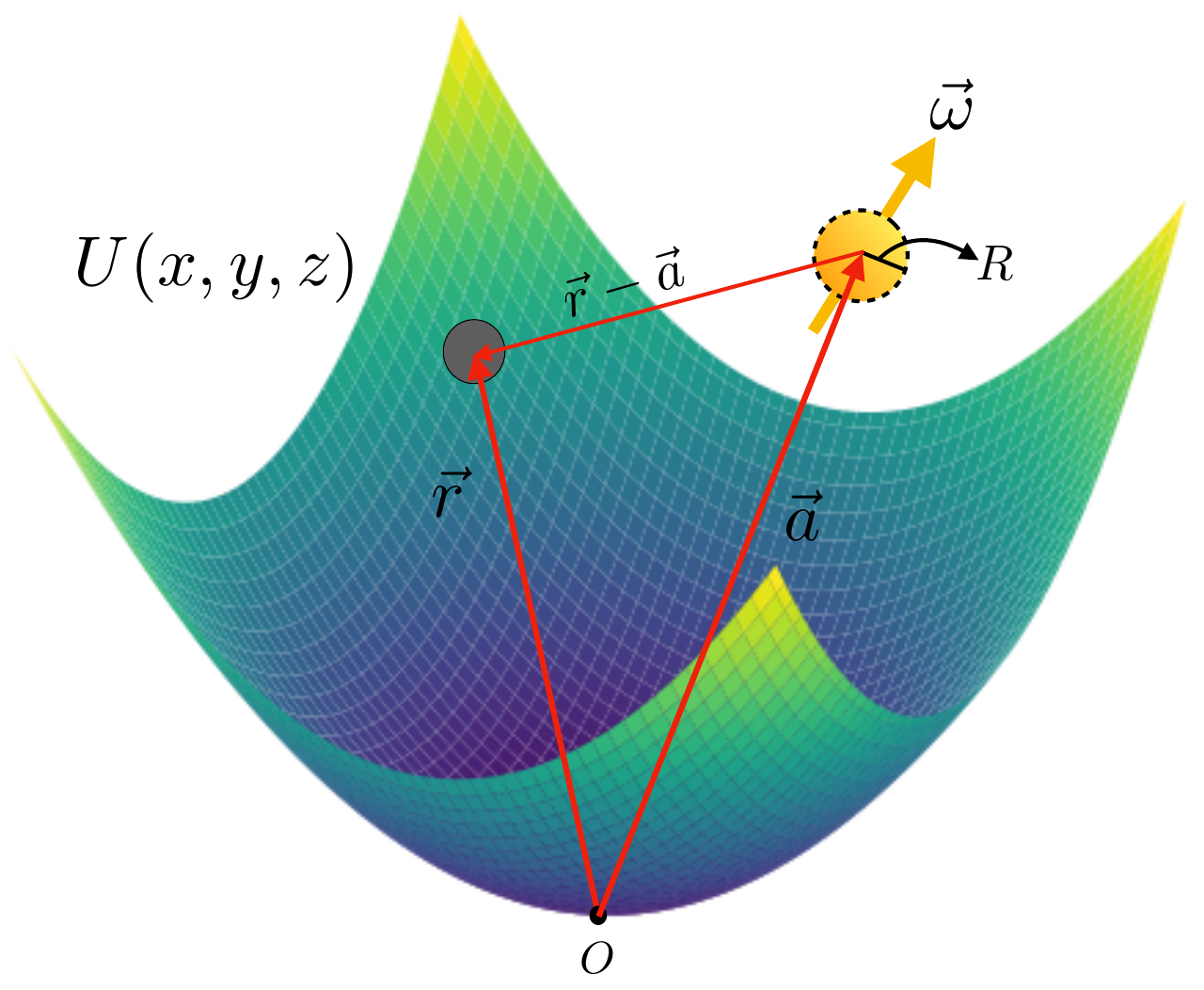}
		\caption{Schematic diagram of a Brownian particle in a 3D anisotropic harmonic trap which is centered at the origin $O$. The flow in the system is generated by a spinor particle of radius $R$ that is spinning with angular velocity $\bm{\omega}$ located at $\bm{a}$.}
	\label{fig:spinor_fig}
\end{figure}

Instead two dimensional shear flow, next we consider a three  dimensional Stokes flow generated around a spinning micro-sphere in highly viscous fluid \cite{dhont1996introduction}. It is motivated by tethered or effectively tethered \cite{dominick2018rotating} spherical bacterium that rotates spontaneously in a highly viscous medium.  We consider the spherical Brownian particle of radius $R$ in three dimensions (3D) at $\bm r=(x,y,z)$ and trapped in a 3D anisotropic harmonic trap
\begin{align}
    U(x,y,z)=\frac{1}{2}[k_x(t)x^2+k_y(t)y^2+k_z(t)z^2],
    \label{eq:potential spinor}
\end{align}
 with time dependent stiffness parameters. Another spherical particle of radius $R$, which we call the ``spinor'', is located at $\bm a=(a_x,a_y,a_z)$ and is spinning with a constant angular velocity $\bm \omega = (\omega_x,\omega_y,\omega_z)$. This creates a viscous flow
 \begin{align}
     \bm v= \frac{\bm \omega\times(\bm r-\bm a)}{|\bm r-\bm a|^3}R^3,
     \label{eq:viscous flow}
 \end{align}
thereby affecting the extraction of work from the system. Such a spinning particle can be used to model the motion of systems like the rotating Janus particle \cite{Mano2017}. In this subsection, we use the simplified model where the spinning axis maintains its direction during the course of the experiment. We refer to  the induced flow thereof as a ``spinor flow''. A schematic diagram of the system is presented in Fig.~\ref{fig:spinor_fig}. 

\begin{figure*}[!t]
	\centering
		\includegraphics[width=\textwidth]{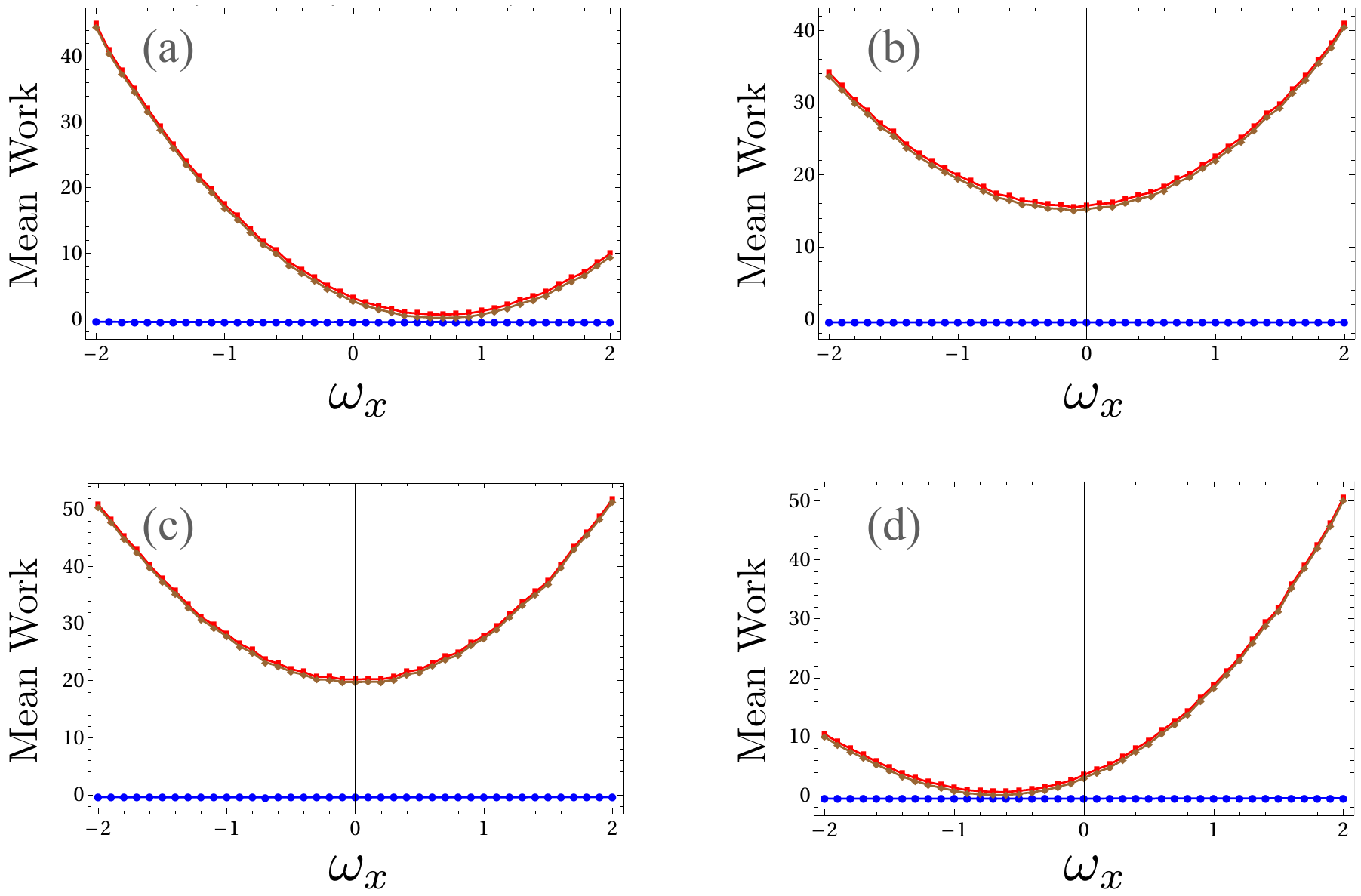}
		\caption{Mean work is plotted as function of $\omega_x$ for different values of $\omega_y$ and $\omega_z$. (a) $\omega_y=\omega_z=1$, (b) $\omega_y=1, \omega_z=-1$, (c) $\omega_y=-1, \omega_z=1$ and (d) $\omega_y=\omega_z=-1$. The values of the stiffness of the external potential has been set to $k_{x,0}=1, k_{y,0}=2$ and $k_{z,0}=3$. The blue dots corresponds to the Jarzynski work, red dots corresponds to the work done by the flow generated in the fluid by the spinor and the brown dots denotes the total work.
   }
	\label{fig:spinor_mean_work_vs_omega_x}
\end{figure*}

Note that if the particle is trapped in an isotropic harmonic trap then the work done by the flow is exactly zero. 
This is because the mean work done by the flow is an average over $\bm{v}\cdot\bm\nabla U$. For isotropic trap, $k_x(t)=k_y(t)=k_z(t)=k(t)$, the above term becomes $k(t)(\bm{\omega}\times (\bm r - \bm a))\cdot \bm r$. Since $(\bm{\omega}\times\bm r)\cdot \bm r$ is identically zero, the only contribution comes from the term $(\bm{\omega}\times\bm a)\cdot \bm r$. Given that the flow fields can assume all possible angles with the constant vector $\bm{\omega}\times\bm a$ with equal probability (see Fig. \ref{fig:toy_model_results1}(a)--(c)), the quantity $\langle(\bm{\omega}\times\bm a)\cdot \bm r\rangle$ vanishes. This prevents the flow field from doing any work when the trap is isotropic.
Hence we consider the trap to be anisotropic here: $k_x(0)$, $k_y(0)$ and $k_z(0)$ are unequal (see caption of Fig. \ref{fig:spinor_mean_work_vs_omega_x}). We also consider here for simplicity that the time dependence of $k_x,~k_y$ and $k_z$ are all linear in time with same periodicity $\tau$.   


We now consider the system of interest to be in proximity to the spinning particle, as described above, and simulate it to study its thermodynamics. 

In Fig.~\ref{fig:spinor_mean_work_vs_omega_x}, we have plotted $\langle W_{\rm tot}\rangle$ as a function of $\omega_x$ for different values of $\omega_y$ and $\omega_z$.
We find that for hyperbolic flows (Figs. (a) and (d)), the mean work done on the system shows a minimum, where its value becomes vanishingly small. In elliptic flows, however, substantial work is done even at the minima. This implies that for certain combinations $(\omega_x,\omega_y,\omega_z)$,  maximum work can be extracted from the system. The blue lines show the values of  $\langle W_{\text{JE}}\rangle$ as a function of $\omega_x$, which are clearly very small when compared to the total work. In fact, the red line for total average work almost coincide with that for the flow-generated work.

\section{Conclusion}\label{conclusion}

Stochastic heat engines have been the subject of intense research. In a typical passive stochastic heat engine, the baths are in equilibrium. The equilibrium thermal fluctuations of the baths are used by the working substance of the engine to produce thermodynamic work {e.g.. \cite{bechinger_2012}}. In case of active stochastic heat engines, baths are out of equilibrium due to the presence of live, motile elements, such as bacteria {e.g. \cite{krishnamurthy2016micrometre}}. These active elements produce non-equilibrium fluctuations in the bath, which are used by the working substance of the engine to produce thermodynamic work. However, these active elements not only responsible for non-equilibrium fluctuations but they also produce hydrodynamic flows in the bath. In general, the baths used to drive the active stochastic engines, can be out of equilibrium due to the presence of hydrodynamic flows. Here we have explored the effects of such flows on the thermodynamics of the stochastic heat engine. 

Here we have considered two different flows - (i) a two dimensional shear flow and (ii) a three dimensional flow generated by a particle spinning in a highly viscous, otherwise quiescent infinite fluid. The working substance is a colloidal bead. It is confined in a harmonic trap that breaths in synchrony with the temperature of the bath that switches between two different values time-periodically, such that work can be produced. However, we find that whether  the system can actually produce work or not, crucially depends on the flow. For both the flows, there are certain regimes in the parameter space of the system, where work can be extracted (i.e. the system can behave as engine) whereas for other regimes it cannot be extracted. This happens because in this case the total work is a sum of two parts : (i) the work due to the breathing harmonic trap (the average of this part is negative which implies that it can be extracted from the system) and (ii) the work injected to the system by the hydrodynamic flow, the average of which is always positive. These two parts compete with each other, and for the parameter space where (i) dominates over (ii), the total work becomes negative. In this case, according to our convention, the work is being extracted from the system and hence the system works as an engine. However, for the parameter space where (ii) dominates over (i), total work becomes positive and consequently the system is not working as an engine.

Realizing the important effects of the flows present in the bath on the thermodynamics of the system, one can proceed to explore the effects of various complex, hydrodynamic flows produced by different types of micro-swimmers present in the active bath \cite{elgeti2015}. The flows and the non-equilibrium fluctuations (both generated by the microswimmers suspended in the non-equilibrium bath) together should provide a more complete description of active bath used in active stochastic heat engines.

\begin{acknowledgments}
P.S.P. gratefully acknowledges research support from Korea Institute for Advanced Study through individual KIAS Grant No.~PG085601. A.S. acknowledges the Core Research Grant (CRG/2019/001492) from DST, Government of India and CY Initiative of Excellence (grant “Investissements d’Avenir” ANR-16-IDEX-0008)
\end{acknowledgments}






\begin{appendix}
\renewcommand{\theequation}{A\arabic{equation}}
\renewcommand{\thefigure}{A\arabic{figure}}
\setcounter{equation}{0}
\setcounter{figure}{0}



\section{Decoupling of coupled Langevin dynamics and its solution}
\label{sec:decoupling}
In the expansion step ($0\le t\le \tau/2$), the whole system is in contact with hot bath at temperature $T_h$. The coupled Langevin equations (in the $(x,y)$ coordinates) are:
\begin{align}
\gamma\dot x&=-k_{\rm e}(t)x+\gamma\omega_x y+\sqrt{2D_{\rm e}}\eta_x, \label{ceom_exp_x}\\
\gamma\dot y&=-k_{\rm e}(t)y+\gamma\omega_y x+\sqrt{2D_{\rm e}}\eta_y,\label{ceom_exp_y}
\end{align}
where $D_{\rm e}=\gamma k_BT_{\rm e}$. In matrix form, the above equations can be rewritten as:
\begin{align}
\frac{d\bm r}{dt}=-\hat\Gamma_{\rm e}(t).\bm r+\hat D_{\rm e}.\bm \eta,
\label{ceom_exp_mat}
\end{align}
where
\begin{align}
\bm r&=\begin{bmatrix} x & y\end{bmatrix}^T; \hspace{0.3cm}
\bm \eta=\begin{bmatrix} \eta_x & \eta_y\end{bmatrix}^T; \nn\\
\hat\Gamma_{\rm e}(t)&=\frac{1}{\gamma}\begin{bmatrix}k_{\rm e}(t) & -\gamma\omega_x\\-\gamma\omega_y & k_{\rm e}(t)\end{bmatrix}; \hspace{0.3cm}
\hat D_{\rm e}=\frac{\sqrt{2D_{\rm e}}}{\gamma}\mathbb{1}_2. \nn
\end{align}
Since the Langevin equation of motions are coupled we need to transform to a new set of coordinates in which $\hat\Gamma(t)$ is diagonal. Let $\hat P$ be the diagonalising matrix that characterises this similarity transformation. Multiplying both sides of Eq.\ref{ceom_exp_mat} by $\hat P^{-1}$ and denoting the transformed coordinates and noises as $\bm{r'}(t)=\hat P^{-1}\cdot \bm r(t)$ and $\bm{\eta'}(t)=\hat P^{-1}\cdot \bm \eta(t)$ respectively, we have
\begin{align}
\frac{d \bm{r'}(t)}{dt}&=-(\hat P^{-1}.\hat\Gamma(t).\hat P).\bm{r'}(t)+(\hat P^{-1}.\hat D_{\rm e}.\hat P).\bm{\eta'}(t)\nn\\
&=-\hat\Gamma'(t).\bm {r'}(t)+\hat D'_{\rm e}.\bm{\eta'}(t).
\label{deom_exp_mat_1}
\end{align}
Here
\begin{align}
\hat P&=\begin{bmatrix}-\omega &\omega\\1&1\end{bmatrix};\hat P^{-1}=\frac{1}{2}\begin{bmatrix}-\omega^{-1}&1\\ \omega^{-1}&1\end{bmatrix};\nn\\
\hat\Gamma'_{\rm e}(t)&=\frac{1}{\gamma}\begin{bmatrix}k_{\rm e}(t)+\gamma\sqrt{\omega_x\omega_y}&0\\0&k_{\rm e}(t)-\gamma\sqrt{\omega_x\omega_y}\end{bmatrix}\nn\\&=\begin{bmatrix}\lambda_x^e(t)&0\\0&\lambda_y^e(t) \end{bmatrix};\nn\\ 
\hat D'_{\rm e}&=\hat P^{-1}.\hat D_{\rm e}.\hat P=\hat D_{\rm e}; \nn\\
\bm{\eta'}(t)&=\frac{1}{2}\begin{bmatrix}-\omega^{-1}\eta_x(t)+\eta_y(t)\\ \omega^{-1}\eta_x(t)+\eta_y(t)\end{bmatrix},\nn
\end{align}
with $\omega=\sqrt{\omega_x/\omega_y}$. The transformed noise has zero mean i.e., $\la\eta'_i(t)\ra=0$, and the noise correlations are given by:
\begin{align}
&\la\eta'_x(t)\eta'_x(t_0)\ra'\nn\\
&=\int\mathcal{D}\bm\eta'P'[\bm\eta']\eta'_x(t)\eta'_x(t_0)\nn\\
&=\frac{1}{4}\int\mathcal{D}\bm\eta P[\bm\eta][-\omega^{-1}\eta_x(t)+\eta_y(t)]\nn\\&\hspace{3cm}\times[-\omega^{-1}\eta_x(t_0)+\eta_y(t_0)]\nn\\
&=\frac{1}{4}\int\mathcal{D}\bm\eta P[\bm\eta][\omega^{-2}\eta_x(t)\eta_x(t_0)+\eta_y(t)\eta_y(t_0)]\nn\\
&=\frac{1}{4}[\omega^{-2}\la\eta_x(t)\eta_x(t_0)\ra+\la\eta_y(t)\eta_y(t_0)\ra]\nn\\
&=\frac{1}{4}(1+\omega^{-2})\delta(t-t_0).
\end{align}
Hence we arrive at the following relations
\begin{align}
\la\eta'_x(t)\eta'_x(t_0)\ra&=\la\eta'_y(t)\eta'_y(t_0)\ra\nn\\
&=\frac{1}{4}(1+\omega^{-2})\delta(t-t_0);\label{etap_corr1}\\
\la\eta'_x(t)\eta'_y(t_0)\ra&=\frac{1}{4}(1-\omega^{-2})\delta(t-t_0)\label{etap_corr2}
\end{align}
Decoupled equation of motion describing the system in the primed coordinates
\begin{align}
\frac{dx'}{dt}&=-\lambda_x^e(t) x'+\frac{\sqrt{2D_{\rm e}}}{\gamma}\eta'_x(t),\label{deom_exp_x}\\
\frac{dy'}{dt}&=-\lambda_y^e(t) y'+\frac{\sqrt{2D_{\rm e}}}{\gamma}\eta'_y(t).\label{deom_exp_y}
\end{align}
Solution of the above two equations are
\begin{align}
x'(t)&=\exp\left[f_{x'}^e(t)\right]\left[x'(0)+\int_{0}^tdt_1F^{\rm e}_{x'}(t_1)\right],\label{sol_exp_x}\\
y'(t)&=\exp\left[f_{y'}^e(t)\right]\left[y'(0)+\int_{0}^tdt_1F^{\rm e}_{y'}(t_1)\right],\label{sol_exp_y}
\end{align}
where $F^{\alpha}_i(t)=\frac{\sqrt{2D_{\alpha}}}{\gamma}\exp\left\{-f_{i}^{\alpha}(t_1)\right\}\eta'_i(t_1)$ with $f_{x'}^e(t)=\frac{[t-(\tau+\tau_2)]^2}{2\tau_1^2}$, $f_{y'}^e(t)=\frac{[t-(\tau-\tau_2)]^2}{2\tau_1^2}$, $\tau_1=\sqrt{\gamma\tau/k_0}$ and $\tau_2=\tau\gamma\sqrt{\omega_x\omega_y}/k_0$.
\\
In the compression step ($\tau/2\le t\le \tau$), the whole system is in contact with cold bath at temperature $T_{\rm c}$. The coupled Langevin equations are:
\begin{align}
\gamma\dot x&=-k_{\rm c}(t)x+\gamma\omega_x y+\sqrt{2D_{\rm c}}\eta_x, \label{ceom_comp_x}\\
\gamma\dot y&=-k_{\rm c}(t)y+\gamma\omega_y x+\sqrt{2D_{\rm c}}\eta_y,\label{ceom_comp_y}
\end{align}
where $D_{\rm c}=\gamma k_BT_{\rm c}$ and $k_{\rm c}(t)=k_0t/\tau$. In a similar way as done before, one can decouple the equation of motion as 
\begin{align}
\frac{dx'}{dt}&=-\lambda_x^c(t) x'+\frac{\sqrt{2D_{\rm c}}}{\gamma}\eta'_x(t) \label{deom_comp_x}\\
\frac{dy'}{dt}&=-\lambda_y^c(t) y'+\frac{\sqrt{2D_{\rm c}}}{\gamma}\eta'_y(t),\label{deom_comp_y}
\end{align}
with $\lambda_x^c(t)=\frac{1}{\gamma}[k_{\rm c}(t)+\gamma\sqrt{\omega_x\omega_y}]$ and $\lambda_y^c(t)=\frac{1}{\gamma}[k_{\rm c}(t)-\gamma\sqrt{\omega_x\omega_y}]$. The solution of the above two equations are
\begin{align}
x'(t)&=\exp\left[-f_{x'}^c(t)\right]\left[x'(\tau/2)+\int_{\tau/2}^tdt_1F^{\rm c}_{x'}(t_1)\right],
\label{sol_comp_x}\\
y'(t)&=\exp\left[-f_{y'}^c(t)\right]\left[y'(\tau/2)+\int_{\tau/2}^tdt_1F^{\rm c}_{y'}(t_1)\right],
\label{sol_comp_y}
\end{align}
where $f_{x'}^c(t)=\frac{(t+\tau_2)^2}{2\tau_1^2}$ and $f_{y'}^c(t)=\frac{(t-\tau_2)^2}{2\tau_1^2}$.
\widetext
\renewcommand{\theequation}{B\arabic{equation}}
\renewcommand{\thefigure}{B\arabic{figure}}
\setcounter{equation}{0}
\setcounter{figure}{0}

\section{Second Moments and crosscorrelation in the primed frame of reference}
\label{sec:SecondMoments}
Using the solutions of the primed coordinates Eqs. \ref{sol_exp_x} and \ref{sol_exp_y} in the expansion step, one can calculate the second moments and the cross-correlation of the coordinates of the primed coordinates as 
\begin{align}
\sigma^e_{x'}(t)&=\la x'^2(t)\ra=\exp[2f_{x'}^e(t)]\left[\la x'^2(0)\ra+\frac{D_{\rm e}\tau_1\sqrt{\pi}(1+\omega^{-2})}{4\gamma^2}\left\{\text{Erf}\left(\frac{\tau+\tau_2}{\tau_1}\right)+\text{Erf}\left(\frac{t-\tau-\tau_2}{\tau_1}\right)\right\}\right],\\
\sigma^e_{y'}(t)&=\la y'^2(t)\ra=\exp[2f_{y'}^e(t)]\left[\la y'^2(0)\ra+\frac{D_{\rm e}\tau_1\sqrt{\pi}(1+\omega^{-2})}{4\gamma^2}\left\{\text{Erf}\left(\frac{\tau-\tau_2}{\tau_1}\right)+\text{Erf}\left(\frac{t-\tau+\tau_2}{\tau_1}\right)\right\}\right],\\
\sigma^e_{x'y'}(t)&=\la x'(t)y'(t)\ra=\exp\left[f_{x'}^e(t)+f_{y'}^e(t)\right]\bigg[\la x'(0)y'(0)\ra+\frac{D_{\rm e}\tau_1\sqrt{\pi}(1-\omega^{-2})}{4\gamma^2}\exp\left(-\frac{\tau_2^2}{\tau_1^2}\right)\nn\\
&\hspace{8.3 cm}\times\left\{\text{Erf}\left(\frac{t-\tau}{\tau_1}\right)+\text{Erf}\left(\frac{\tau}{\tau_1}\right)\right\}\bigg].
\label{eq:variance expansion}
\end{align} 
Using the solutions of the primed coordinates Eqs. \ref{sol_comp_x} and \ref{sol_comp_y} in the compression step, one can calculate the second moments and the cross-correlation of the coordinates of the primed coordinates as 
\begin{align}
\sigma^c_{x'}(t)&=\la x'^2(t)\ra=\exp[-2f_{x'}^c(t)]\left[\la x'^2(\tau/2)\ra+\frac{D_{\rm c}\tau_1\sqrt{\pi}(1+\omega^{-2})}{4\gamma^2}\left\{\text{Erfc}\left(\frac{t+\tau_2}{\tau_1}\right)-\text{Erfc}\left(\frac{\tau/2+\tau_2}{\tau_1}\right)\right\}\right],\\
\sigma^c_{y'}(t)&=\la y'^2(t)\ra=\exp[-2f_{y'}^c(t)]\left[\la y'^2(\tau/2)\ra+\frac{D_{\rm c}\tau_1\sqrt{\pi}(1+\omega^{-2})}{4\gamma^2}\left\{\text{Erfc}\left(\frac{t-\tau_2}{\tau_1}\right)-\text{Erfc}\left(\frac{\tau/2-\tau_2}{\tau_1}\right)\right\}\right],\\
\sigma^c_{x'y'}(t)&=\la x'(t)y'(t)\ra=\exp[-f_{x'}^c(t)-f_{y'}^c(t)]\bigg[\la x'(\tau/2)y'(\tau/2)\ra+\frac{D_{\rm c}\tau_1\sqrt{\pi}(1-\omega^{-2})}{4\gamma^2}\exp\left(\frac{\tau_2^2}{\tau_1^2}\right)\nn\\
&\hspace{9 cm}\times\left\{\text{Erfc}\left(\frac{t}{\tau_1}\right)-\text{Erfc}\left(\frac{\tau}{2\tau_1}\right)\right\}\bigg].
\label{eq:variance compression}
\end{align} 
\renewcommand{\theequation}{C\arabic{equation}}
\renewcommand{\thefigure}{C\arabic{figure}}
\setcounter{equation}{0}
\setcounter{figure}{0}
\section{Calculation of average work in non-quasistatic and quasistatic regime}
\label{sec:AverageWorkCalculation}
In the non-quasistatic regime, we use the expressions of the second moments and cross-correlations in the previous section to calculate the average of the rate of work done on the system. 
Average rate of work done in the expansion step:
\begin{align}
\la\dot w_{\rm e}(t)\ra&=\la D_tU(x,y,t)\ra=\la [\partial_t+\bm v.\bm {\nabla}]U(x,y,t)\ra=\la \partial_t U+\omega_x y\partial_xU+\omega_y x\partial_yU \ra,\nn\\
&=\frac{1}{2}\dot k_{\rm e}(t)[\la x^2(t)\ra+\la y^2(t)\ra]+(\omega_x+\omega_y)k_{\rm e}(t)\la x(t)y(t)\ra,\nn\\
&=\frac{1}{2}\dot k_{\rm e}(t)[(1+\omega^2)\{\sigma^e_{x'}(t)+\sigma^e_{y'}(t)\}+2(1-\omega^2)\sigma^e_{x'y'}(t)]+\omega(\omega_x+\omega_y)k_{\rm e}(t)\{\sigma^e_{y'}(t)-\sigma^e_{x'}(t)\},
\label{wdot_exp}
\end{align}
where we use the transformation rules between the primed and unprimed frames of reference to obtain the third equality:
\begin{align}
    \la x^2(t)\ra&=\int dxP(x) x^2(t)=\int dxdyP_{\rm tot}(x,y) x^2(t)\nn\\
    &=\int dx'dy'P'_{\rm tot}(x',y')\omega^2(x'-y')^2\nn\\
    &=\omega^2\int dx'dy'P'_{\rm tot}(x',y')(x'^2+y'^2-2x'y')\nn\\
    &=\omega^2[\la x'^2\ra'+\la y'^2\ra'-2\la x'y'\ra']=\omega^2[\sigma^e_{x'}(t)+\sigma^e_{y'}(t)-2\sigma^e_{x'y'}(t)].
\end{align}
Hence we arrive at the following relations
\begin{align}
    \la x^2(t)\ra&=\la \{\omega[-x'(t)+y'(t)]\}^2\ra=\omega^2[\sigma^e_{x'}(t)+\sigma^e_{y'}(t)-2\sigma^e_{x'y'}(t)]\nn\\
    \la y^2(t)\ra&=\la [x'(t)+y'(t)]^2\ra=\sigma^e_{x'}(t)+\sigma^e_{y'}(t)+2\sigma^e_{x'y'}(t)\nn\\
    \la x(t)y(t)\ra&=\la\omega[-x'(t)+y'(t)][x'(t)+y'(t)]\ra=\omega[\sigma^e_{y'}(t)-\sigma^e_{x'}(t)].
\end{align}
Average rate of work done in the compression step:
\begin{align}
\la\dot w_{\rm c}(t)\ra&=\la D_tU(x,y,t)\ra=\la [\partial_t+\bm v.\bm {\nabla}]U(x,y,t)\ra=\la \partial_t U+\omega_x y\partial_xU+\omega_y x\partial_yU \ra,\nn\\
&=\frac{1}{2}\dot k_{\rm c}(t)[\la x^2(t)\ra+\la y^2(t)\ra]+(\omega_x+\omega_y)k_{\rm c}(t)\la x(t)y(t)\ra,\nn\\
&=\frac{1}{2}\dot k_{\rm c}(t)[(1+\omega^2)\{\sigma^c_{x'}(t)+\sigma^c_{y'}(t)\}+2(1-\omega^2)\sigma^c_{x'y'}(t)]+\omega(\omega_x+\omega_y)k_{\rm c}(t)\{\sigma^c_{y'}(t)-\sigma^c_{x'}(t)\},
\label{wdot_comp}
\end{align}
where, as in the expansion case, the transformation rules between the primed and unprimed frames have been used:
\begin{align}
    \la x^2(t)\ra&=\la \{\omega[-x'(t)+y'(t)]\}^2\ra=\omega^2[\sigma^c_{x'}(t)+\sigma^c_{y'}(t)-2\sigma^c_{x'y'}(t)]\nn\\
    \la y^2(t)\ra&=\la [x'(t)+y'(t)]^2\ra=\sigma^c_{x'}(t)+\sigma^c_{y'}(t)+2\sigma^c_{x'y'}(t)\nn\\
    \la x(t)y(t)\ra&=\la\omega[-x'(t)+y'(t)][x'(t)+y'(t)]\ra=\omega[\sigma^c_{y'}(t)-\sigma^c_{x'}(t)].
\end{align}

Average of the total work done in the full cycle is
\begin{align}
\la W_{\text{tot}}\ra&=\int_0^{\tau}\left\la\frac{d w_{\text{tot}}(t)}{dt}\right\ra dt =\int_0^{\tau/2}\la\dot w_{\rm e}(t)\ra dt+\int_{\tau/2}^{\tau}\la\dot w_{\rm c}(t)\ra dt
\end{align}
Using Eqs. \ref{wdot_exp} and \ref{wdot_comp}, one can easily compute the average of total work. Since the expression is cumbersome, we avoid it and instead provide a figure depicting the behaviour of the average of the total work on the rotating parameters, i.e., $\omega_x$ and $\omega_y$ Fig.~\ref{fig:toy_model_results1}.

\paragraph*{\textbf{Quasistatic limit:}} In the next part we calculate the average of the total work in the quasistatic limit. Multipying Eq. \ref{deom_exp_x} with $x'$ and averaging over all trajectories we get
\begin{align}
\frac{1}{2}\D{\sigma^e_{x'}}{t}=-\lambda_x^e(t)\sigma^e_{x'}+\frac{\sqrt{2D_{\rm e}}}{\gamma}\la x'(t)\eta_x'(t)\ra,
\end{align}
where $\sigma^e_{x'}(t)=\la x'^2(t)\ra$ in the expansion step when the system is connected to the hot bath. The average in the second term of the above equation is calculated using Eq. \ref{sol_exp_x} as follows:
\begin{align}
\la x'(t)\eta_x'(t)\ra&=\exp\bigg[f_{x'}^e(t)\bigg]\bigg[x'(0)\la \eta_x'(t)\ra+\frac{\sqrt{2D_{\rm e}}}{\gamma}\int_0^tdt_1\la\eta_x'(t_1)\eta_x'(t)\ra\exp\bigg\{-f_{x'}^e(t_1)\bigg\}\bigg],\nn\\
&=\exp\bigg[f_{x'}^e(t)\bigg]\frac{\sqrt{2D_{\rm e}}}{\gamma}\frac{(1+\omega^{-2})}{4}\int_0^tdt_1\delta(t-t_1)\exp\bigg\{-f_{x'}^e(t_1)\bigg\},\nn\\
&=\frac{\sqrt{2D_{\rm e}}(1+\omega^{-2})}{4\gamma},
\end{align}
where in the second equality we have used Eq. \ref{etap_corr1}. Therefore, the dynamical equation of the second moment of $x'(t)$ is given by
\begin{align}
\D{\sigma^e_{x'}}{t}=-2\lambda_x^e(t)\sigma^e_{x'}+\frac{D_{\rm e}}{\gamma^2}(1+\omega^{-2}).
\label{psigma1}
\end{align}
Similarly, one can find the dynamical equation for the second moment of $y'(t)$:
\begin{align}
\D{\sigma^e_{y'}}{t}=-2\lambda_y^e(t)\sigma^e_{y'}+\frac{D_{\rm e}}{\gamma^2}(1+\omega^{-2}).
\label{psigma2}
\end{align}
The dynamical equation of $\sigma^e_{x'y'}(t)=\la x'(t)y'(t)\ra$ is obtained by multiplying Eq. \ref{deom_exp_x} by $y'$ and Eq. \ref{deom_exp_y} by $x'$ and adding both the equations:
\begin{align}
y'\D{x'}{t}+x'\D{y'}{t}=-[\lambda_x^e(t)+\lambda_y^e(t)]x'y'+\frac{\sqrt{2D_{\rm e}}}{\gamma}[ y'(t)\eta_x'(t)+ x'(t)\eta_y'(t)].
\end{align} 
Averaging both sides with respect to the ensemble of trajectories, we obtain
\begin{align}
\D{\sigma^e_{x'y'}}{t}=-[\lambda_x^e(t)+\lambda_y^e(t)]\sigma^e_{x'y'}+\frac{\sqrt{2D_{\rm e}}}{\gamma}[\la y'(t)\eta_x'(t)\ra+ \la x'(t)\eta_y'(t)\ra].
\end{align}
The averages in the second term of the above are calculated using Eq. \ref{sol_exp_x} and Eq. \ref{sol_exp_y} as follows:
\begin{align}
\la y'(t)\eta_x'(t)\ra&=\exp\bigg[f_{y'}^e(t)\bigg]\bigg[y'(0)\la\eta_x'(t)\ra+\frac{\sqrt{2D_{\rm e}}}{\gamma}\int_0^tdt_1\la\eta_y'(t_1)\eta_x'(t)\ra\exp\bigg\{-f_{y'}^e(t_1)\bigg\}\bigg]\nn\\
&=\exp\bigg[f_{y'}^e(t)\bigg]\frac{\sqrt{2D_{\rm e}}(1-\omega^{-2})}{4\gamma}\int_0^tdt_1\delta(t-t_1)\exp\bigg\{-f_{y'}^e(t)\bigg\}\nn\\
&=\frac{\sqrt{2D_{\rm e}}(1-\omega^{-2})}{4\gamma}.
\end{align}
Similarly, we have 
\begin{align}
\la x'(t)\eta_y'(t)\ra=\frac{\sqrt{2D_{\rm e}}(1-\omega^{-2})}{4\gamma}.
\end{align}
Therefore, the dynamical equation of $\sigma^h_{x'y'}(t)$ is given by
\begin{align}
\D{\sigma^e_{x'y'}}{t}=-[\lambda_x(t)+\lambda_y(t)]\sigma^e_{x'y'}+\frac{D_{\rm e}}{\gamma^2}(1-\omega^{-2}).
\label{psigma12}
\end{align} 
The quasistatic limit is obtained by $\tau\rightarrow\infty$ while $t/\tau$ is finite. We consider a new variable (a rescaled time variable) $s=t/\tau$. This implies that $\D{}{t}=\D{s}{t}\D{}{s}=\frac{1}{\tau}\D{}{s}$. Therefore, changing the variable, Eqs. \ref{psigma1},\ref{psigma2},\ref{psigma12} become:
\begin{align}
\frac{1}{\tau}\D{\sigma^{e,q}_{x'}}{s}&=-2\lambda_x^e(s\tau)\sigma^{e,q}_{x'}+\frac{D_{\rm e}}{\gamma^2}(1+\omega^{-2}),\nn\\
\frac{1}{\tau}\D{\sigma^{e,q}_{y'}}{s}&=-2\lambda_y^e(s\tau)\sigma^{e,q}_{y'}+\frac{D_{\rm e}}{\gamma^2}(1+\omega^{-2}),\nn\\
\frac{1}{\tau}\D{\sigma^{e,q}_{x'y'}}{s}&=-[\lambda_x^e(s\tau)+\lambda_y^e(s\tau)]\sigma^{e,q}_{x'y'}+\frac{D_{\rm e}}{\gamma^2}(1-\omega^{-2}).
\end{align}
Here, $\sigma^q_{i}$ is the quasistatic values of $\sigma_{i}$. In the quasistatic limit ($\tau\rightarrow\infty$) we can drop the L.H.S of the above equations as it tends to zero and hence we obtain the quasistatic values of second moments and crosscorrrelation:
\begin{align}
\sigma^{e,q}_{x'}(s)&=\frac{D_{\rm e}(1+\omega^{-2})}{2\gamma^2}\frac{1}{\lambda_x^e(s\tau)}=\frac{D_{\rm e}(1+\omega^{-2})}{2\gamma}\frac{1}{k_0(1-s)+\gamma\sqrt{\omega_x\omega_y}},\nonumber\\
\sigma^{e,q}_{y'}(s)&=\frac{D_{\rm e}(1+\omega^{-2})}{2\gamma^2}\frac{1}{\lambda_y^e(s\tau)}=\frac{D_{\rm e}(1+\omega^{-2})}{2\gamma}\frac{1}{k_0(1-s)-\gamma\sqrt{\omega_x\omega_y}},\nonumber\\
\sigma^{e,q}_{x'y'}(s)&=\frac{D_{\rm e}(1-\omega^{-2})}{\gamma^2}\frac{1}{\lambda_x^e(s\tau)+\lambda_y^e(s\tau)}=\frac{D_{\rm e}(1-\omega^{-2})}{2\gamma}\frac{1}{k_0(1-s)}.
\label{eq:sigma_h primes}
\end{align}
Using new variable $s$, we can rewrite the expression (Eq. \ref{wdot_exp}) of average rate of work done in the expansion step as 
\begin{align}
\left\la\frac{d w_{\rm e}(s)}{ds}\right\ra&=-\frac{k_0}{2}\omega[(1+\omega^2)\{\sigma^e_{x'}(s)+\sigma^e_{y'}(s)\}+2(1-\omega^2)\sigma^e_{x'y'}(s)]\nn\\
&\hspace{4 cm}-\tau\omega(\omega_x+\omega_y)k_0\left(1-s\right)[\sigma^e_{x'}(s)-\sigma^e_{y'}(s)].
\label{eq:WexpExact}
\end{align} 
In the quasistatic limit, $\tau\rightarrow\infty$ and hence the last term will dominate. Therefore the average rate of work in the expansion process in quasistatic limit is given by
\begin{align}
\left\la\frac{d w^q_{\rm e}(s)}{ds}\right\ra&=-\tau\omega(\omega_x+\omega_y)k_0\left(1-s\right)[\sigma^{e,q}_{x'}(s)-\sigma^{e,q}_{y'}(s)]\nn\\
&=-\tau\omega(\omega_x+\omega_y)k_0(1-s)\frac{D_{\rm e}(1+\omega^{-2})}{2\gamma}\bigg[\frac{1}{k_0(1-s)+\gamma\sqrt{\omega_x\omega_y}}-\frac{1}{k_0(1-s)-\gamma\sqrt{\omega_x\omega_y}}\bigg]\nn\\
&=-\frac{\tau D_{\rm e}\omega(\omega_x+\omega_y)^2}{2\gamma\omega_x}k_0(1-s)\frac{(-2\gamma\sqrt{\omega_x\omega_y})}{k_0^2(1-s)^2-\gamma^2\omega_x\omega_y}\nn\\
&=\tau D_{\rm e}(\omega_x+\omega_y)^2\frac{k_0(1-s)}{k_0^2(1-s)^2-\gamma^2\omega_x\omega_y}
\label{eq:WexpExact}
\end{align}
In the compression step, the only difference is $D_{\rm e}$ is now $D_{\rm c}=\gamma k_BT_{\rm c}$ and $k(s)=k_{\rm c}(s)=k_0s$. Therefore, the relevant expressions in the compression step is
\begin{align}
\sigma^{c,q}_{x'}(s)&=\frac{D_{\rm c}(1+\omega^{-2})}{2\gamma^2}\frac{1}{\lambda_x^c(s\tau)}=\frac{D_{\rm c}(1+\omega^{-2})}{2\gamma}\frac{1}{k_0s+\gamma\sqrt{\omega_x\omega_y}},\nonumber\\
\sigma^{c,q}_{y'}(s)&=\frac{D_{\rm c}(1+\omega^{-2})}{2\gamma^2}\frac{1}{\lambda_y^c(s\tau)}=\frac{D_{\rm c}(1+\omega^{-2})}{2\gamma}\frac{1}{k_0s-\gamma\sqrt{\omega_x\omega_y}},\nonumber\\
\sigma^{c,q}_{x'y'}(s)&=\frac{D_{\rm c}(1-\omega^{-2})}{\gamma^2}\frac{1}{\lambda_x^c(s\tau)+\lambda_y^c(s\tau)}=\frac{D_{\rm c}(1-\omega^{-2})}{2\gamma}\frac{1}{k_0s}.
\label{eq:sigma_c primes}
\end{align}
Using new variable $`s'$, we can rewrite the expression (Eq. \ref{wdot_comp}) of average rate of work done in the compression step as 
\begin{align}
\left\la\frac{d w_{\rm c}(s)}{ds}\right\ra
&=\frac{k_0}{2}[(1+\omega^2)\{\sigma^c_{x'}(s)+\sigma^c_{y'}(s)\}+2(1-\omega^2)\sigma^c_{x'y'}(s)]\nn\\
&\hspace{4 cm}-\tau\omega(\omega_x+\omega_y)k_0s\{\sigma^c_{x'}(s)-\sigma^c_{y'}(s)\}.
\label{eq:WcomExact}
\end{align}
In the quasistatic limit (i.e., large $\tau$ limit), the average rate of work in the compression process is given by
\begin{align}
\left\la\frac{d w^q_{\rm c}(s)}{ds}\right\ra
&=-\tau\omega(\omega_x+\omega_y)k_0s\{\sigma^{c,q}_{x'}(s)-\sigma^{c,q}_{y'}(s)\}\nn\\
&=-\tau\omega(\omega_x+\omega_y)k_0s\frac{D_{\rm c}(1+\omega^{-2})}{2\gamma}\bigg[\frac{1}{k_0s+\gamma\sqrt{\omega_x\omega_y}}-\frac{1}{k_0s-\gamma\sqrt{\omega_x\omega_y}}\bigg]\nn\\
&=-\frac{\tau D_{\rm c}\omega(\omega_x+\omega_y)^2}{2\gamma\omega_x}k_0s\frac{(-2\gamma\sqrt{\omega_x\omega_y})}{k_0^2s^2-\gamma^2\omega_x\omega_y}\nn\\
&=\tau D_{\rm c}(\omega_x+\omega_y)^2\frac{k_0s}{k_0^2s^2-\gamma^2\omega_x\omega_y}
\label{eq:WcomApprox}
\end{align}
The total average work in the quasistatic limit is
\begin{align}
\la W^q_{\text{tot}}\ra&=\int_0^1\left\la\frac{d w^q_{\text{tot}}(s)}{ds}\right\ra ds\nn\\
&=\int_0^{1/2}\left\la\frac{d w^q_{\rm e}(s)}{ds}\right\ra ds+\int_{1/2}^1\left\la\frac{d w^q_{\rm c}(s)}{ds}\right\ra ds\nn\\
&=\tau D_{\rm e}(\omega_x+\omega_y)^2\int_0^{1/2}\frac{k_0(1-s)ds}{k_0^2(1-s)^2-\gamma^2\omega_x\omega_y}+\tau D_{\rm c}(\omega_x+\omega_y)^2\int_{1/2}^1\frac{k_0sds}{k_0^2s^2-\gamma^2\omega_x\omega_y}\nn\\
&=\tau D_{\rm e}(\omega_x+\omega_y)^2\bigg[\frac{1}{2k_0}\ln\bigg\{\frac{4(k_0^2-\gamma^2\omega_x\omega_y)}{k_0^2-4\gamma^2\omega_x\omega_y}\bigg\}\bigg]+\tau D_{\rm c}(\omega_x+\omega_y)^2\bigg[\frac{1}{2k_0}\ln\bigg\{\frac{4(k_0^2-\gamma^2\omega_x\omega_y)}{k_0^2-4\gamma^2\omega_x\omega_y}\bigg\}\bigg]\nn\\
&=\frac{\tau (D_{\rm e}+D_{\rm e})(\omega_x+\omega_y)^2}{2k_0}\ln\bigg[\frac{4(k_0^2-\gamma^2\omega_x\omega_y)}{k_0^2-4\gamma^2\omega_x\omega_y}\bigg].
\label{sq_tot_work}
\end{align}

\renewcommand{\theequation}{D\arabic{equation}}
\renewcommand{\thefigure}{D\arabic{figure}}
\setcounter{equation}{0}
\setcounter{figure}{0}
\section{Expressions of work in Spinor model}
\label{app:spinor}
In case of spinor model, the the flow is generated due to the presence of a spinning Brownian particle of radius $R$ at a position $\bm a$. The particle is spinning  with an angular velocity given by $\bm{\omega}$ keeping its axis of rotation fixed. The expression of the flow at any point $\bm r$ generated by it is given by
\begin{align}
    \bm v(\bm r)=\frac{R^3[\bm{\omega}\times(\bm r-\bm a)]}{|\bm r-\bm a|^3}.
\end{align}
The system consists of a Brownian particle moving in three dimension in presence of an external harmonic potential $U(x,y,z,t)=\frac{1}{2}[k_x(t)x^2+k_y(t)y^2+k_z(t)z^2]$ with time varying stiffness. The time variation of the potential stiffness is given by
\begin{align}
    k_i(t)&=k_{i,\rm e}(t)=k_{i,0}\bigg(1-\frac{t}{\tau}\bigg)\hspace{2 cm} 0\leq t\leq \frac{1}{2}\tau\nn\\
    &=k_{i,\rm c}(t)=k_{i,0}\frac{t}{\tau}\hspace{3 cm} \frac{1}{2}\tau\leq t\leq \tau,
\end{align}
for $i=x,y,z$. It is to be noted that $k_{x,0}\neq k_{y,0}\neq k_{z,0}$.
Average rate of work done in the expansion step:
\begin{align}
\la\dot w_{\rm e}(t)\ra&=\la D_tU(x,y,z,t)\ra=\la [\partial_t+\bm v(\bm r).\bm {\nabla}]U(x,y,z,t)\ra=\la \partial_t U(x,y,z,t)\ra+\la(\bm v(\bm r).\bm {\nabla})U(x,y,z,t) \ra,\nn\\
&=\frac{1}{2}[\dot k_{x,\rm e}(t)\la x^2\ra+\dot k_{y,\rm e}(t)\la y^2\ra+\dot k_{z,\rm e}(t)\la z^2\ra]+\frac{R^3}{|\bm r-\bm a|^3}\bigg[k_{x,\rm e}(t)\la[\omega_y(z-a_z)-\omega_z(y-a_y)]x\ra\nn\\
&\hspace{1cm}+k_{y,\rm e}(t)\la[\omega_z(x-a_x)-\omega_x(z-a_z)]y\ra+k_{z,\rm e}(t)\la[\omega_x(y-a_y)-\omega_y(x-a_x)]z\ra\bigg],\nn\\
&=\frac{1}{2}[\dot k_{x,\rm e}(t)\sigma_x(t)+\dot k_{y,\rm e}(t)\sigma_y(t)+\dot k_{z,\rm e}(t)\sigma_z(t)]+\frac{R^3}{|\bm r-\bm a|^3}\bigg[k_{x,\rm e}(t)[\omega_y\la zx\ra-\omega_z \la xy\ra-(\omega_y a_z-\omega_z a_y)\la x\ra]\nn\\
&\hspace{1 cm}+k_{y,\rm e}(t)[\omega_z\la xy\ra-\omega_x\la yz\ra-(\omega_za_x-\omega_xa_z)\la y \ra]+k_{z,\rm e}(t)[\omega_x\la yz\ra-\omega_y\la zx\ra-(\omega_xa_y-\omega_ya_x)\la z \ra]\bigg],\nn\\
&=\frac{1}{2}[\dot k_{x,\rm e}(t)\sigma_x(t)+\dot k_{y,\rm e}(t)\sigma_y(t)+\dot k_{z,\rm e}(t)\sigma_z(t)]\nn\\
&\hspace{1cm}+\frac{R^3}{|\bm r-\bm a|^3}\bigg[\omega_z\la xy\ra\{k_{y,\rm e}(t)-k_{x,\rm e}(t)\}+\omega_x\la yz\ra\{k_{z,\rm e}(t)-k_{y,\rm e}(t)\}+\omega_y\la zx\ra\{k_{x,\rm e}(t)-k_{z,\rm e}(t)\}\nn\\
&\hspace{2.75 cm}
-k_{x,\rm e}(t)(\omega_y a_z-\omega_z a_y)\la x\ra-k_{y,\rm e}(t)(\omega_za_x-\omega_xa_z)\la y \ra-k_{z,\rm e}(t)(\omega_xa_y-\omega_ya_x)\la z \ra\bigg].
\label{wdot_exp_spinor}
\end{align}
Similarly, the average rate of work done in the compression step:
\begin{align}
\la\dot w_{\rm c}(t)\ra&=\frac{1}{2}[\dot k_{x,\rm c}(t)\sigma_x(t)+\dot k_{y,\rm c}(t)\sigma_y(t)+\dot k_{z,\rm c}(t)\sigma_z(t)]\nn\\
&\hspace{1 cm}+\frac{R^3}{|\bm r-\bm a|^3}\bigg[\omega_z\la xy\ra\{k_{y,\rm c}(t)-k_{x,\rm c}(t)\}+\omega_x\la yz\ra\{k_{z,\rm c}(t)-k_{y,\rm c}(t)\}+\omega_y\la zx\ra\{k_{x,\rm c}(t)-k_{z,\rm c}(t)\}\nn\\
&\hspace{2.75 cm}-k_{x,\rm c}(t)(\omega_y a_z-\omega_z a_y)\la x\ra-k_{y,\rm c}(t)(\omega_za_x-\omega_xa_z)\la y \ra-k_{z,\rm c}(t)(\omega_xa_y-\omega_ya_x)\la z \ra\bigg].
\label{wdot_comp_spinor}
\end{align}
\end{appendix}

\bibliographystyle{unsrt}
\bibliography{draft/draft2_19.06.24}


\end{document}